\documentclass{article}
\usepackage{geometry}                
\usepackage{graphicx}
\usepackage{amssymb}
\usepackage{epstopdf}
\usepackage{latexsym}
\usepackage{dcolumn}
\usepackage{epsfig}
\title{Light Sterile Neutrinos, Lepton Number Violating Interactions and the LSND Neutrino Anomaly}
\author{{K. S. Babu}\\
 {\it Department of Physics, Oklahoma State University, Stillwater, OK 70478}
\and
Douglas W. McKay\\
{\it Department of Physics and Astronomy, University of Kansas, Lawrence, KS 66045}
\and
Irina Mocioiu\\ 
{\it Department of Physics, The Pennsylvania State University, University Park, PA 16802}
\and
Sandip Pakvasa\\
 {\it Department of Physics and Astronomy, University of Hawaii, Honolulu, HI 96822}}
\date{\today}

\begin{document}
\maketitle

\begin{abstract}
We develop the consequences of introducing a purely leptonic, lepton number violating non-standard interaction (NSI) and standard model neutrino mixing with a fourth, sterile neutrino in the analysis of short-baseline, neutrino experiments.  We focus on the muon decay at rest (DAR) results from the Liquid Scintillation Neutrino Experiment (LSND) and the Karlsruhe and Rutherford Medium Energy Neutrino Experiment (KARMEN).  We make a comprehensive analysis of lepton number violating, NSI effective operators and find nine that affect muon decay relevant to LSND results. Two of these  preserve the standard model (SM) value 3/4 for the Michel $\rho$ and $\delta$ parameters and, overall, show favorable agreement with precision data and the $\bar{\nu}_{e}$ signal from LSND data. We display theoretical models that lead to these two effective operators. In the model we  choose to apply to DAR data, both $\overline{\nu}_{e}$ appearance from $\overline{\nu}_{\mu}$ oscillation and $\overline{\nu}_{e}$ survival after production from NSI  decay of the $\mu^{+}$ contribute to the expected signal. This is a unique feature of our scheme. We find a range of parameters where  both  experiments can be accommodated consistently with recent global, sterile neutrino fits to short baseline data.  We comment on  implications of the models for new physics searches at colliders and comment on further implications of the lepton number violating interactions plus sterile neutrino-standard neutrino mixing.
\end{abstract}
\thispagestyle{empty}
\newpage
\section{Introduction}

The early positive results reported by the 30 m baseline LSND $\overline{\nu}_{e}$ appearance search \cite{LSND1} were hard to reconcile with then-current results from "long baseline" solar neutrino and atmospheric neutrino limits on mixing and oscillation parameters \cite{solarandatmo}. This naturally led to conjectures about new physics that incorporates both short and long baseline data, invoking, for example, new interactions \cite{jm,bands}, sterile neutrinos \cite{barger}, \cite{conrad}, extra dimensions \cite{dlp} and quantum decoherence \cite{bandm}.

Among the approaches, two that have gained considerable interest are new interactions and new oscillations induced by one or more additional, super-weakly interacting, or sterile, neutrinos, with recent global fits to short-baseline appearance and disappearance data provided in \cite{kopp}, \cite{glll2}, \cite{ggms}, \cite{gariazzo} and to disappearance data alone \cite{glll1}.  There continues to be a great deal of interest in the possibility of sterile neutrinos \cite{SK15}, \cite{T2K15}, \cite{IC}, including applications in astrophysics and cosmology \cite{whitepaper}, with some indications that, under certain assumptions, cosmology constraints may cause a problem \cite{bggns}, \cite{noconcord}.
 Theoretical consistency requirements plus tight limits on $\mu$ and $\tau$ branching fractions made lepton flavor-violating NSI models unworkable \cite{bandg99}.\footnote{A model-independent lepton-flavor violating NSI setup with one or more sterile neutrinos can give a good global description of short baseline data \cite{aands}, but it considers neither lepton number violation nor consistency with charged lepton flavor constraints.}. However, as pointed out in \cite{bandg99},  $\Delta$L $\neq$ 0 interactions can evade the charged lepton flavor constraints. 
 Even so, high precision measurements of the Michel parameters in $\mu$ decay still provide strong constraints on the form of any proposed $\Delta$L $\neq$ 0 effective Lagrangian \cite{bands}.\footnote{In Ref.\cite{bands}, two specific models were presented (denoted $(B_{2})_1$ and $(B_2)_{2}$ in our Table 2 in Sec.(2.2)). Both of these models have the feature that the anomalous muon decay that explains the LSND signal would also lead to shifts in all of the Michel parameters in $\mu$ decay by an amount characterized by the LSND signal strength. Results from the TWIST collaboration \cite{twist11, twist12} have since excluded these options.} 
 
In this paper, we undertake a comprehensive analysis of lepton number violating muon decay, and find that among possible effective operators there are two that retain the SM prediction of $\rho=\delta=\frac{3}{4}$,   We develop theoretical models that lead to these two self-consistent effective operators.  We need the oscillations provided by a sterile neutrino in addition, since a model with \emph{only} $\mu^+ \rightarrow e^{+}$ + $\overline{\nu}_{\mu}$+$\overline{\nu}_{e}$ and no oscillations as the DAR explanation of the LSND signal would directly conflict with the absence of a signal in KARMEN data. With oscillations, the difference in baselines, 30 m for LSND and 17 m for KARMEN, allows some leeway in fitting both signals.
Moreover,  MiniBoone \cite{mb} would have seen no indication of an appearance signal, since their source is the semileptonic decay $\pi^{-} \rightarrow \mu^{-}$ +$\overline{\nu}_{\mu}$, which requires a subsequent oscillation to $\overline{\nu}_{e}$ to produce an appearance signal. \emph{Only} the muon DAR experiments are sensitive to the $\Delta$L$ \neq$ 0 new interactions we consider here. Nonetheless, we find that our fitted  oscillation parameters are compatible with those of global fits that include the MiniBoone and other data which rely on neutrinos from semileptonic decays.

With these considerations and LSND \cite{LSND2} and KARMEN \cite{KARMEN} in mind as the two experiments with muon DAR as their $\bar{\nu}$ source, we classify a set of nine possible models with $\Delta$L $\neq$ 0, retaining those that are consistent with current experimental constraints on muon decay parameters and lepton decay branching ratios.  We then add an additional, sterile neutrino that mixes with the three standard model neutrinos. The model we choose to study in detail has the striking feature that \emph{both} antineutrinos in the flux from the source can produce the "appearance" signal: the $\overline{\nu}_{e}$s that come from oscillations of both standard model (SM) and NSI decays of the $\mu^{+}$ and the $\overline{\nu}_{e}$s that don't disappear after being directly produced in the NSI decay mode of the $\mu^{+}$. We develop the models, summarize their properties and pursue the consequences of the most promising example for LSND and KARMEN in the following four sections. We outline further research directions, summarize and conclude.

\section{Survey of anomalous $\mu$-decay models}

The effective Lagrangian we want must produce a muon decay relevant to LSND that results in a final state with a net lepton number different from that of the decaying muon, is consistent with the SU(2)$_{L}\times$U(1) symmetry of the SM, and satisfies the experimental constraints on the Michel decay parameters of the muon \cite{kando}, which have been measured to high precision \cite{twist11, twist12, PDG14}.  The SM shows overall agreement with the measured values.

At the level of the muon decay process, we consider models with $\nu_{eL}$, $\mu_{L}$ or $\mu_{R}$, $e_{L}$ or $e_{R}$, and any flavor of active neutrino with  $\nu_{aL}$ or it's conjugate and models with a singlet, two-component sterile neutrino, $\nu_{R}$. We define three categories determined by the choice of neutrino: 
\begin{eqnarray*}
 A: \nu&=&\nu_{aL},\; \mathrm{active \;left\!-\!handed\; neutrino,}  \\ 
 B: \nu&=&(\nu_{a}^{C})_{R}, \;\mathrm{active\; antineutrino,}\\ 
 C: \nu&=&\nu_{R},\; \mathrm{singlet\;right\!-\!handed\; sterile\;neutrino}.
 \label{eq:categories}
\end{eqnarray*}
Within each category, one can choose right or left helicity projections or charge conjugation projections of $\mu$, $e$ or $\nu$, and within these projections there are generally several ways to choose the SU(2)$_{L}$ multiplet contraction of indices to form a scalar and consequent effective $\mu$-decay 4-Fermion operator.

\subsection{Models with lepton number conserved but lepton flavor variable }

To illustrate category A, we review the lepton-flavor violating but lepton-number conserving cases considered by Bergmann and Grossman \cite{bandg99}. Effective  4-Fermion operators responsible for the flavor-violating $\mu$-decay can be written in category A, with choice of $e$ and $\mu$ as left-handed SU(2)$_{L}$ doublet members, in Lorentz vector and scalar forms, or similarly with $e$ and $\mu$ as right-handed SU(2)$_{L}$ singlets. In the doublet forms, we have  
\begin{eqnarray}
({\cal L}_{A.1})_{1}&=&(\bar{\mu}_{L}\gamma_{\lambda}e_{L})(\bar{\nu}_{aL}\gamma^{\lambda}\nu_{eL})+H.c.\\ 
({\cal L}_{A.1})_{2}&=&(\nu_{eL}^{T}Ce_{L})(\mu_{L}^{T}C\nu_{aL})^*+H.c.,
\label{eq:bgVS}
\end{eqnarray}
where index $a$ can be $e$, $\mu$, or $\tau$ and $C$ is the charge conjugation operator. Both forms lead to the decay $\mu^{+} \rightarrow \bar{\nu_e}+e^{+}+\nu_{a}$, with lepton number preserved but flavor violated, directly providing a source of $\bar{\nu}_{e}$s to produce an $e^{+}$ signal by inverse $\beta$-decay in the detector. 
SU(2)$_{L}\times$U(1) invariant effective Lagrangians that contain Eq.(1) and Eq.(2) follow by replacing each field by its corresponding SU(2)$_{L}$ doublet, $\psi_{i}^T= (\nu_{i},\ell_{i})$:
\begin{eqnarray}
({\cal L}_{A.1})_{1}&\rightarrow&(\bar{\psi}_{\mu L}\gamma_{\lambda}\psi_{eL})(\bar{\psi}_{aL}\gamma^{\lambda}\psi_{eL})+H.c.\\ 
({\cal L}_{A.1})_{2}&\rightarrow&(\psi_{eL}^{T}C\psi_{eL})(\psi_{\mu L}^{T}C\psi_{aL})^*+H.c.,
\label{eq:bgVsu2}
\end{eqnarray} 
and then contracting the doublet indices to make an overall SU(2)$_{L}\times$U(1) singlet $L_{eff}$. Regardless of the choice of SU(2)$_{L}\times$U(1) construction, for the V-A form one ends up with a term $(\bar{\mu}_{L}\gamma_{\lambda}e_{L})(\bar{\ell}_{aL}\gamma^{\lambda}e_{L})$ + ..., which gives a contribution to $\mu^{+}\rightarrow e^{+}+e^{+}+e^{-}$ when $a= e$, a contribution to $e^{-}+\mu^{+}\rightarrow e^{+}+\mu^{-}$ when $a=\mu$ and a contribution to $\tau^{+}\rightarrow e^{+}+e^{+}+\mu^{-}$, when $a=\tau$.
All of these processes are so tightly constrained by experiment \cite{PDG14}, that the bound on the new physics coefficient rules out the chance that it could explain the LSND anomaly.  The same situation results from the S form, where the term $(e_{L}^{T}Ce_{L})(\mu_{L}^{T}C\ell_{aL})^*$ + ...contributes to the same processes as the V-A form for the corresponding flavor choice for the index $a$.  Under the A-category \ref{eq:categories}, one can replace the left-handed doublet members $e_{L}$ and $\mu_{L}$  by right-handed singlet fields and perform the same analysis as above and find the same severe constraints. SU(2)$_{L}\times$U(1)$_{Y}$ breaking effects, introduced through the vacuum expectation value of the Higgs field in four-fermion operators with higher dimensions, cannot be significant enough to change this conclusion.  \emph{The lepton number conserving but flavor violating new physics contributions to muon decay cannot account for the LSND anomaly} \cite{bandg99}.

\subsection{Models with neither lepton number nor flavor conserved}

In category B, the outgoing antineutrinos and positron in $\mu^{+}$ decay carry lepton number -3 and change the lepton number by 2 units.  A purely charged leptonic decay of $\mu$ or $\tau$ consistent with SU(2)$_{L}\times$U(1) cannot be constructed without violating electric charge conservation, so the limits on "conventional" rare decays discussed above don't apply.  The two lepton helicity combinations in this category are  $\mu_{R}$ and $e_{L}$ (B.1) and $\mu_{L}$ and $e_{R}$ (B.2). In these first cases, the four-Fermion effective interactions we want for the $\Delta$L=2 $\mu$-decay modes $\mu^{+}\rightarrow e^{+}+\bar{\nu}_{e}+\bar{\nu}_{a}$ are 
\begin{eqnarray}
{\cal L}_{B.1}&=&(\bar{\mu}_{R}e_{L})((\overline{\nu_{a}^{C}})_{R}\nu_{eL})\; \mathrm{and}\; (\bar{\mu}_{R}\nu_{eL})((\overline{\nu_{a}^{C}})_{R}e_{L}),\; \mathrm{while}\\ 
{\cal L}_{B.2}&=&(\bar{\mu}_{L}\gamma_{\lambda}\nu_{eL}) ((\overline{\nu_{a}^{C}})_{R}\gamma^{\lambda}e_{R}) \;\mathrm{and}\; (\bar{\mu}_{L}e_{R})((\overline{\nu_{a}^{C}})_{R}\nu_{eL}),
\label{eq:b1b2}
\end{eqnarray}
where the Hermitian conjugate terms will be understood from here on. With a scalar Higgs doublet of the standard model, $H^{T}=(H^{+},H^{0})$, we can write an SU(2)$_{L}$$\times$U(1) singlet effective Lagrangian for the first case in three ways, namely
\begin{equation}
({\cal L}_{B.1})_{1,2,3}=(\bar{\mu}_{R}\psi_{eL}^{i})(\psi_{aL}^{Tj}C\psi_{eL}^{k})H^{l}\;\{\;(1)\;\varepsilon_{ij}\varepsilon_{kl},\;(2)\;\varepsilon_{ik}\varepsilon_{jl}\;\ \mathrm{and}\;(3)\;\varepsilon_{il}\varepsilon_{jk}\}, 
\label{B1cases}
\end{equation}
where $i, j, k,$ and $l$ are indices of the components of the SU(2)$_{L}$ doublets and the epsilon symbols are the constant SU(2)$_{L}$ antisymmetric two-index tensors.

To display the effective 4-Fermion operators relevant to $\mu$-decay, we break out the terms proportional to the vacuum expectation value of the neutral component of the Higgs field:
\begin{eqnarray}
({\cal L}_{B.1})_{1}&\rightarrow&[(\bar{\mu}_{R}\nu_{eL})(\ell_{aL}^{T}C\nu_{eL})-(\bar{\mu}_{R}e_{L})(\nu_{aL}^{T}C\nu_{eL})]\langle |H^{0}|\rangle\; ,\\ 
({\cal L}_{B.1})_{2}&\rightarrow&[(\bar{\mu}_{R}\nu_{eL})(\nu_{aL}^{T}Ce_{L})-(\bar{\mu}_{R}e_{L})(\nu_{aL}^{T}C\nu_{eL})]\langle |H^{0}|\rangle\;\mathrm{and}\\ 
({\cal L}_{B.1})_{3}&\rightarrow&[(\bar{\mu}_{R}\nu_{eL})(\nu_{aL}^{T}Ce_{L})-(\bar{\mu}_{R}\nu_{eL})(\ell_{aL}^{T}C\nu_{eL})]\langle |H^{0}|\rangle.
\end{eqnarray}
The second terms in $(L_{B.1})_{1}$ and $(L_{B.1})_{2}$ are the same, while $(L_{B.2})_{2}$ and $(L_{B.3})_{3}$ share their first terms. Both terms of each case must be included to ensure SU(2)$_{L}\times$U(1) invariance, of course. 
The other member of this $\Delta$L=2 category follows from exchanging the helicity labels on $\mu$ and e.  The basic 4-Fermion SU(2)$_{L}\times$U(1) invariant forms are then
\begin{eqnarray}
({\cal L}_{B.2})_{1}&=&(\bar{\psi}^{i}_{\mu L} \gamma_{\lambda}\psi_{eLj})(\psi_{aLk}^{T}C\gamma^{\lambda}e_{R})H_{i}H_{l}H_{m}\epsilon^{jl}\epsilon^{km}\rightarrow[(\bar{\mu}_{L}\gamma_{\lambda}\nu_{eL})(\nu_{aL}^{T}C\gamma^{\lambda}e_{R})]\langle |H^{0}|\rangle^{3}, \\ 
({\cal L}_{B.2})_{2}&=&(\bar{\psi}_{\mu L}^{i}e_{R})(\psi_{aLj}^{T}C\psi_{eLk})H_{i}H_{l}H_{m}\epsilon^{jl}\epsilon^{km}\rightarrow[(\bar{\mu}_{L}e_{R})(\nu_{aL}^{T}C\nu_{eL})]\langle|H^{0}|\rangle^{3},
\label{eq:b2s}
\end{eqnarray}
where the 4-Fermion piece of each invariant form that is proportional to $\langle |H^{0}|\rangle^{3}$ is indicated by the arrows.

Category C introduces a new, right-handed, SU(2)$_{L}$ singlet neutrino that does not mix with the standard model neutrinos and interacts only through its NSI \footnote{In Sec. 3.2 we introduce a left-handed, sterile neutrino that mixes with the three neutrinos of the standard model and implements new oscillation effects driven by its O(1 eV$^{2}$) mass.}. It contributes a super-weak term to the effective Lagrangian that produces a direct $\mu$-decay mode $\mu^{+}\rightarrow e^{+}+\bar{\nu}_{e}+\nu_{R}$ . 
The $\mu$-decay forms read
\begin{eqnarray}
{\cal L}_{C.1}&=&(\bar{\mu}_{L}e_{R})(\bar{\nu}_{R}\nu_{eL})\; \mathrm{and}\; (\bar{\mu}_{L}\gamma_{\lambda}\nu_{eL})(\bar{\nu}_{R}\gamma^{\lambda}e_{R}), \; \mathrm{while}  \\ 
{\cal L}_{C.2}&=&(\bar{\mu}_{R}\nu_{eL})(\bar{\nu}_{R}e_{L}),\;(\bar{\mu}_{R}e_{L})(\bar{\nu}_{R}\nu_{eL}) \; \mathrm{and}\; (e_{L}^{T}C\nu_{eL})(\mu_{R}^{T}C\nu_{R})^{*}.
\label{eq:c1c2}
\end{eqnarray}
The SU(2)$_{L}$$\times$U(1) forms from which these effective 4-Fermion Lagrangians arise through coupling to the Higgs doublet are
\begin{equation}
({\cal L}_{C.1})_{1}=(\bar{\psi}_{\mu L}^{i}e_{R})(\bar{\nu}_{R}\psi_{eLj})H_{i}H_{k}\varepsilon^{jk}\rightarrow (\bar{\mu}_{L}e_{R})(\bar{\nu}_{R}\nu_{eL})\langle |H^{0}|\rangle^{2},
\label{eq:c1-1}
\end{equation}
for the first term of Eq. (13), and
\begin{equation}
({\cal L}_{C.1})_{2}=(\bar{\psi}_{\mu L}^{i}\gamma_{\lambda}\psi_{eLj})(\bar{\nu}_{R}\gamma^{\lambda}e_{R})H_{i}H_{k}\varepsilon^{jk}\rightarrow (\bar{\mu}_{L}\gamma_{\lambda}\nu_{eL})(\bar{\nu}_{R}\gamma^{\lambda}e_{R})\langle |H^{0}|\rangle^{2},
\label{eq:c1-2}
\end{equation}
for the second. Again the arrows indicate the result of extracting the effective 4-Fermion pieces of the invariant interactions.
The second version of this singlet, right-handed neutrino category exchanges the R and L labels on $\mu_{L}$ and $e_{R}$, and the SU(2)$_{L}\times$U(1) invariant forms are
\begin{equation}
({\cal L}_{C.2})_{1}=(\bar{\mu}_{R}\psi_{eL}^{i})(\bar{\nu}_R\psi_{eL}^{j})\varepsilon_{ij}\rightarrow(\bar{\mu}_{R}\nu_{eL})(\bar{\nu}_{R}e_{L})-(\bar{\mu}_{R}e_{L})(\bar{\nu}_{R}\nu_{eL}) 
\label{c2-1}
\end{equation}
which covers the first two terms in Eq.(14), and

\begin{equation}
({\cal L}_{C.2})_{2}=(\psi_{eL}^{Ti}C\psi_{eL}^{j})(\mu_{R}^{T}C\nu_{R})^{*}H^{k}H_{i}^{*}\varepsilon _{jk}\rightarrow (e_{L}^{T}C\nu_{eL})(\mu_{R}^{T}C\nu_{R})^{*}\langle |H^{0}|\rangle^{2},
\label{eq:c2-2}
\end{equation}
which covers the third.

This completes our summary of the effective Lagrangian models with SU(2)$_{L}\times$U(1) symmetry with relevance to $\mu$-decay. One can cast the operators in different forms by making Fierz transformations on the operators we have presented, but the physics content remains the same.  We devote the remainder of this section to testing each of the lepton number violating cases for consistency with precision Michel parameter data \cite{twist11, twist12, PDG14}. 

\subsection{Study of the models' contributions to the $\mu$-decay Michel-parameters}
As presented in the preceding discussion, there are nine distinct variants of the $\Delta$L $\neq$ 0 NSI scheme we are studying.  If no explicit breaking of the SU(2)$_{L}\times$U(1) symmetry at the effective Lagrangian level is included, only the generic parameter $\epsilon$ that governs the strength of the NSI contribution to the LSND signal needs to be considered. We assume this to be the case for the present discussion. The definition of $\epsilon$ in the case (L$_{B.1})_{3}$ that we study in Sec. 3, for example, is given by ${\cal L}_{eff}=\frac{4G_{F}}{\surd 2}\left[(\bar{e}_{L}\gamma_{\lambda}\nu_{eL})(\bar{\nu}_{\mu L}\gamma^{\lambda}\mu_{L})+2\epsilon(\bar{\mu}_{R}\nu_{eL})(\nu_{\mu L}^{T}{C}e_{L})\right]$.  The factor 2 in the definition of $\epsilon$ compensates for the factor 1/4 in the decay rate with the S,P structures compared to the V-A structure. When $\epsilon$=1, the full contribution to the rate results as required for consistent normalization.  The results for the general cases are illustrated in the Appendix. 

The Lorentz and helicity structures of the effective 4-Fermion interactions we have introduced are generally quite different from those of the SM, so we expect that there will be differences in the prediction for the energy and angular distributions of the $e^{+}$ in the $\mu$-decay final state compared to that predicted by the SM.  Because the linear terms in $\epsilon$ are proportional to the neutrino masses and are completely negligible, the deviations of Michel parameters from their SM values are at order $|\epsilon^{2}|$.  Directly computing the complete spin-averaged decay distribution in the final state electron energy for each model would give the answer for each case, or adapting the exhaustive studies of $\mu$ decay parameterizations in the literature to the particular cases of interest here can yield the results we want \cite{fgj, kando}. We rely primarily on the latter approach.

The $\chi^{2}$ that tests the simultaneous fit of the Michel parameters and the LSND $\bar{\nu}_{e}$ rate for each model as a function of $\epsilon$ can be written
\begin{equation}
\chi^{2}(\epsilon)=(\frac{\rho_{twist}-\rho(\epsilon)}{\sigma(\rho)_{twist}})^{2}+(\frac{\delta_{twist}-\delta(\epsilon)}{\sigma(\delta)_{ twist}})^{2}+(\frac{\xi_{twist}-\xi(\epsilon)}{\sigma(\xi)_{ twist}})^{2}+(\frac{P_{lsnd}-|\epsilon|^{2}}{\sigma(P)_{lsnd}})^{2},
\label{eq:chisq}
\end{equation}
where $\rho_{twist}$ is the numerical value of the TWIST global fit for $\rho$ and $\sigma(\rho)$ is the 90\% C.L. uncertainty in the fit value and similarly for the $\delta$ and $\xi$ values and their uncertainties and for those of the LSND oscillation probability.\footnote{The 90\%C.L. value for $\sigma(P)_{lsnd}$ is taken to be $\surd2.71\times\sigma_{expt}$ =0.0013, which includes both statistical and systematic errors.} The expressions for $\rho(\epsilon)$, $\delta(\epsilon)$ and $\xi(\epsilon)$, the TWIST values from their global fit given in their Table VII \cite{twist12}, and the best fit and its corresponding $\chi^{2}$ per degree of freedom, are listed below the respective models at the head of each column in Table 1. The standard model values for $\rho$, $\delta$ and $\xi$ are 3/4, 3/4 and 1, as shown in the column headed "SM".\footnote{The $\eta$ parameter is zero in the SM and in all of our nine $\Delta$L$\neq 0$ models. The measurement of $\eta$ is consistent with zero within 1 $\sigma$, but it is an order of magnitude less precise than the others and is not included.}   The $\chi^{2}$ per degree of freedom for the SM refers only to the comparison to the Michel parameter values in the TWIST global fit. It is not the $\epsilon$=0 fit to the Michel plus LSND data.  If one adds the last (LSND) term to the SM evaluation of Eq. (\ref{eq:chisq}) and divides by 4, the SM $\chi^{2}$ per degree of freedom is 3.34, a full unit higher than our best fit candidate models $(B_{1})_{3}$ and 
$(C_{2})_{2}$. Surprisingly, the average contribution of the Michel parameters and that of the LSND signal to the SM $\chi^2$ are comparable. 

\begin{table}[ht] 
\begin{center}
   \caption{The multiplicative corrections to the SM values of the Michel parameters in each of the 9 $\Delta$L $\neq$ 0 model variants we consider is given in the column below the corresponding model's name. The third column under "TWIST" lists the 90\%/C.L. global fit value for each parameter, as given in Table VII of Ref.\cite{twist12}.  The best fit value of $\epsilon$ and the $\chi$ squared per degree of freedom at 90\% C.L. for the fit are give in the last two items in each column. In the column labeled "SM", the $\chi^2$/3 value comes from the first three terms in Eq. \ref{eq:chisq}, with the SM values for the Michel parameters in their corresponding terms. See text for the discussion of the full SM comparison.} 
     \smallskip
   \begin{tabular}{||l|l|l|l|l|c|l|l|l|c||} 
 \hline 

     Michel & SM&TWIST & $(B_{1})_{1}$  &  $(B_{1})_{2}$  & $(B_{1})_{3}$ & $(B_{2})_{1,2}$ &  $(C_{1})_{1,2}$  & $(C_{2})_{1}$ & $(C_{2})_{2}$  \\  \hline
      
           $\rho(\epsilon)$&0.75 & 0.74960& $1-|\epsilon|^{2}$ & $1-\frac{2}{3}|\epsilon|^{2}$ & 1 &$1-|\epsilon|^{2}$ & $1-|\epsilon|^{2}$ & $1-\frac{2}{3}|\epsilon|^{2}$  &  1  \\ 
                                    &  &$\pm$0.00019 &                       &                                             &    &                                                        &                                                      &      & \\ \hline
           
           $\delta(\epsilon)$ &0.75& 0.74997& $1-3|\epsilon|^2$ & $1-2|\epsilon|^2$ & 1 & $1+3|\epsilon|^2$ & $1+3|\epsilon|^2$ & $1-2|\epsilon|^2$ & 1  \\ 
                                      &   &$\pm$0.00028 & & &   &  &  &  &        \\ \hline
           
           $\xi(\epsilon)$ & 1.0& 0.99897 & $1+2|\epsilon|^2$ & $1+\frac{2}{3}|\epsilon|^2$ & $1-2|\epsilon|^2$ & $1-4|\epsilon|^2$  & $1-4|\epsilon|^2$  & $1+\frac{2}{3}|\epsilon|^2$ &  $1-2|\epsilon|^2$  \\  
                                       &  &$\pm$0.00046 &  &  &  &  &  &  &     \\  \hline  
           
             $\epsilon_{fit}$& & & 0.0063 & 0.012 & 0.024 & 0.012 & 0.012 & 0.012 & 0.024  \\ \hline
           
           $\chi^2$/3& 3.15 & & 6.61 & 6.29 & 2.30 & 3.0 & 3.0 & 6.29 & 2.30 \\ \hline
             
       \end{tabular}
      \end{center}
\label{tab:rho}
\end{table}

To illustrate the application of the set-up of Sec. II.C.1 in Kuno and Okada \cite{kando} to our problem, we derive the entry for the model $({\cal L}_{B1})_{1}$; for example $\rho_{SM}=\frac{3}{4}\rightarrow\frac{3}{4}(1-|\epsilon|^{2})$ in Table 1.  Though Kuno and Okada consider a general flavor violating but lepton number conserving case, we can readily adapt it to our models by casting the Lagrangian terms in the same lepton/neutrino pairing form for each Lorentz and helicity structure to read off the appropriate coefficients from their Eq.(43), which is reproduced in our Appendix as Eq.(\ref{eq:genL}). The effective Michel-$\rho$ parameter is then identified from Eq.(\ref{eq:rho}).  The second term of our Eq.(8) is the relevant piece for this discussion, and we use a Fierz rewriting to put it in the generic form: $(\bar{\mu}_{R}e_{L})(\bar{\nu}_{aR}^{C}\nu_{eL})$=$\frac{1}{2}(\bar{\mu}_{R}\nu_{eL})(\bar{\nu}_{aR}^{C}e_{L})+\frac{1}{8}(\bar{\mu}_{R}\sigma_{\mu \nu}\nu_{eL})(\bar{\nu}_{aR}^{C}\sigma^{\mu \nu}e_{L})$.  The effective Lagrangian reads
\begin{equation}
{\cal L}_{eff}=-\frac{4G_{F}}{\surd 2}\left[(\bar{e}_L\gamma_{\mu}\nu_{eL})(\bar{\nu}_{\mu L}\gamma^{\mu}\mu_{L})+\epsilon(\bar{\mu}_{R}\nu_{eL})(\bar{\nu}_{aR}^Ce_{L})+\frac{\epsilon}{4}(\bar{\mu}_{R}\sigma_{\mu \nu}\nu_{eL})(\bar{\nu}_{aR}^{C}\sigma^{\mu \nu}e_{L})\right].
\end{equation}
In the notation of Kuno and Okada, Eq.(\ref{eq:genL}) in our Appendix, $g_{LR}^{S}=\epsilon^{*}$ and $g_{LR}^{T}=\frac{\epsilon^{*}}{2}$, so the $\rho$ parameter, Eq.(\ref{eq:rho}), reads $\rho=\frac{3}{4}(1-2|g_{LR}^{T}|^{2}-Re(g_{LR}^{S}g_{LR}^{T*}))=\frac{3}{4}(1-|\epsilon|^{2})$, as shown in Table 1. The remaining cases can be evaluated similarly, and we give further discussion in the Appendix.

Among the nine models,  $(B_{1})_3$ and $(C_{2})_{2}$ show the best agreement between the limits on $|\epsilon|$ imposed by the precision measurements of the Michel parameters and the value needed to accommodate the LSND $\bar{\nu}_e$ signal. A substantial amount of the $\chi^{2}$ per degree of freedom comes from the tension between the tree-level values of $\rho$ and $\xi$ and the global fit to data, which amounts to about two "90\%C.L deviations" in each of these cases.  For illustration, we will use models $(B_{1})_{3}$ and $(C_{2})_{2}$, whose best fit values of $|\epsilon|$ are the largest and most promising for our LSND study, to illustrate the construction of renormalizable $\Delta$L$\neq$ 0 models and their implications in the next section.  Notice that their fits improve the overall $\chi^2$ compared to the SM fit, despite the addition of the term representing the LSND appearance signal. We may interpret this as a sign that there is room for new physics in the TWIST global fit to the Michel parameters.

\subsection {Renormalizable Models for anomalous muon decay}

In this section we present briefly two explicit models that generate the anomalous lepton number violating decays of the muon.
At low energies, these models reduce to  effective interactions of the muon that leave both $\rho$ and $\delta$ parameters in muon decay 
at their SM value of $3/4$.  Thus we provide the basic mechanism for generating model $(B_1)_3$ and $(C_2)_2$ of Table 1,
with the effective Lagrangians given in Eqs. (10) and (18) respectively. We shall focus on extensions of the SM with the
introduction of scalar fields.

\vspace*{0.1in}
\noindent{\bf Model \boldmath{$(B_1)_3$:}}  The effective Lagrangian of Eq. (10) can be obtained by the addition of two
scalar fields to the SM spectrum denoted by $\phi(1,2,1/2)$ and $\eta^+(1,1,1)$ where the quantum numbers under
$SU(3)_C \times SU(2)_L \times U(1)_Y$ gauge symmetry are indicated.  
The $\phi = (\phi^+, \phi^0)^T$ field is a second Higgs doublet field,   but the vacuum expectation
value of $\phi^0$ is assumed to be negligible compared to that of the SM field $H^0$.  The new  couplings of these fields contain the terms
\begin{equation}
{\cal L}_{(B_1)_3} = y_1\, (\overline{\mu}_R \psi_{eL}^i) \,\phi^j \,\varepsilon_{ij}  + y_2\, (\psi_{eL}^{Ti} C \psi_{aL}^j) \,\eta^+ \,\varepsilon_{ij} + 
\mu \, H^i \phi^j \eta^- \varepsilon_{ij} + H.c.,
\label{B13}
\end{equation}
where $a=\mu$ or $\tau$.
\begin{figure}[h!]
	\centering
\includegraphics[scale=0.5]{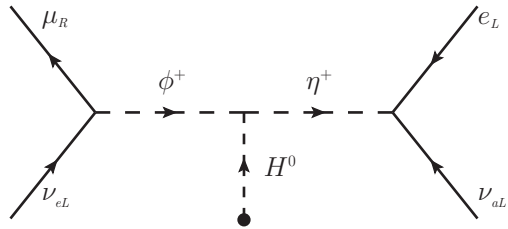}
\vspace*{-0.2in}
\caption{Diagram inducing anomalous muon decay in model $(B_1)_3$.}
\label{diag1}
\end{figure}
\noindent The Feynman diagram shown in Fig. \ref{diag1} will induce the desired $\Delta L = 2$ muon decay $\mu^+ \rightarrow e^+ \overline{\nu}_e \overline{\nu}_a$.  The strength of the interaction relative to the standard decay obtained from this diagrams is
\begin{equation}
\epsilon = \frac{1}{4\sqrt{2}G_{F}} \left(\frac{y_1 y_2 \mu v}{m^2_\phi m^2_\eta}\right)
\label{ep}
\end{equation}
where we have assumed that the $\phi$ and $\eta$ fields are nearly mass eigenstates. As an example, consider the choice $y_1=0.3,\, y_2=1.0,\,
m_\phi = 400$ GeV, $m_\eta = 200$ GeV and $\mu = 400$ GeV.  In this case, $\epsilon = 0.057$, with a corresponding LSND signal strength of
$\epsilon^2 = 3.3 \times 10^{-3}$. When our $\Delta$L $\neq 0$ models are combined with the effects of a sterile neutrino that mixes with the three SM
neutrinos, the requirements on $\epsilon$ weaken considerably, as we show in following sections.  

This explicit realization can be tested in collider experiments.  First, the exchange of the neutral $\phi^0$ scalar would contribute
in the $t$-channel to the cross section for the process $e^+ e^- \rightarrow \mu^+ \mu^-$.  This contribution will modify the
SM prediction at the level of the LSND signal, which is well within current experimental limits.

The model can be tested more directly at the LHC through the pair production of $\phi^+$ and $\phi^0$ scalars.  The relevant processes
are $p p \rightarrow \phi^+ \phi^0$ via $W$ exchange and $p p \rightarrow \phi^0 \overline{\phi^0}$ via $Z$ exchange.  The dominant decay
of $\phi^0$ is into $e^+ \mu^-$ and for $\phi^+$ is into $\mu^+ \nu$.  This would lead to trilepton signature with missing energy as
well as to four-lepton signals.  In each case, the invariant mass of $e^+$ and $\mu^-$ would show a resonant structure, corresponding to
the mass of $\phi^0$.  The production cross section for these processes at the LHC at 13 TeV is in the few fb range (depending on the
mass of $\phi)$.  With more accumulated data, there is a chance of seeing this resonance in the future.  Observe that the masses of
the $\phi^+$ and $\eta^+$ fields cannot be taken to large values, since that would diminish the anomalous muon decay amplitude.  To be
relevant for the LSND anomaly, these scalars should have masses below about 500 GeV, though again our comment on the effect of including
a sterile neutrino applies

Note that the $\phi^0$ field couples primarily to $e$ and $\mu$, and thus carries approximate electron number and muon number.
As a result, it does not lead to any lepton flavor violating decays.  Furthermore, since the charged leptons in the muon decay diagram
are distinct ($e$ and $\mu$), this diagram cannot be closed to generate neutrino masses.  Thus, this model is quite different from the
Zee model \cite{zee} of neutrino masses, where $m_\nu$ is induced by closing diagrams similar to Fig. \ref{diag1}.
We are assuming, as usual, that small neutrino masses arise via the seesaw mechanism involving heavy right-handed neutrino fields.

Since there is a new decay channel for the muon, the Fermi coupling determined in nuclear beta decay will differ slightly from
the coupling determined from $\mu$ decay.  Such a small difference can be interpreted as a shift in the value of $|V_{ud}|$.  Unitarity
of the first row of the Cabibbo-Kobayashi-Maskawa (CKM) matrix would set a limit on any such differences, however, this limit at the level of $3 \times 10^{-3}$
allows some room for consistency with the LSND signal.

\begin{figure}[h!]
	\centering
\includegraphics[scale=0.5]{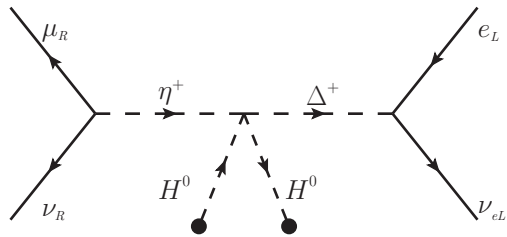}
\vspace*{-0.2in}
\caption{Diagram generating anomalous muon decay in model $(C_2)_2$.}
\label{diag2}
\end{figure}

\vspace*{0.1in}
\noindent{\bf Model \boldmath{$(C_2)_2$:}}  This model includes a sterile neutrino $\nu_R$.  The relevant interaction of Eq. (18) is
obtained by introducing two scalar fields $\Delta(1,3,1)$ and $\eta^+(1,1,1)$.  The diagram of Fig. \ref{diag2} would induce the needed
amplitude for muon decay.  The Lagrangian relevant for the diagram is given as
\begin{equation}
{\cal L}_{(C_2)_2} = y_1 (\psi_{eL}^{T}\, C \,i\tau_2 \tau_a \,\psi_{eL}) \Delta_a 
+ y_2\, (\psi_{eL}^{Ti} C \psi_{aL}^j) \,\eta^+ \,\varepsilon_{ij} + \alpha\, \eta^- H^\dagger \Delta H + H.c.
\end{equation}

As in Eq. (\ref{ep}), we obtain $\epsilon^2 \approx 3 \times 10^{-3}$ for mass parameters of order $500$ GeV and couplings
$y_{1,2} = (0.3 - 1)$.  Again, these scalar particles cannot be made too heavy.  In this model
we assume that the vacuum expectation value of the $\Delta^0$ field is sufficiently small as to be consistent with neutrino masses.
The contribution to $m_\nu$ from heavy right-handed neutrinos is assumed to be dominant. The model can be tested at the LHC in the
pair production of doubly charged scalar $\Delta^{++}$, which has a unique signal of decaying dominantly into $e^+ e^+$.
There are limits on such scalars from ATLAS \cite{ATLAS} and CMS collaborations \cite{CMS}, which set the mass of $\Delta^{++} \geq 500$ GeV
if this particle decays 100\% of the time time into $e^+ e^+$.  We note that within the model the decay $\Delta^{++} \rightarrow
\Delta^+ W^+$ may also occur, which could weaken this limit.

\section{$\mu^{+}$ $\Delta$L=2 decay, $\overline{\nu}_{\mu}$ and $\overline{\nu}_{e}$ fluxes, and $\nu_{s}$-mixing-induced oscillations}

For the remainder of the paper, we pursue the phenomenological consequences of applying our $\Delta$L $\neq$ 0 NSI,  combined with a sterile neutrino,  to LSND and KARMEN data.  As argued in the preceding section, a model that is consistent with SM  gauge symmetry, is safe from rare charged-lepton process like $\mu\rightarrow eee$, is consistent with constraints imposed by precision measurements of Michel parameters and allows the most leeway for values of $|\epsilon|$ is exemplified by the $\Delta$L=2 case $(L_{B1})_{3}$, Eq. (10).  We reproduce its form here in order to make the discussion as self-contained as possible. 
\begin{equation}
(\bar{\mu}_{R}(\psi^{k}_{e})_{L})((\psi^{l}_{a})_{L}^{T}C(\psi^{m}_{e})_{L})H^{n}\varepsilon_{kn}\varepsilon_{lm}\rightarrow (\bar{\mu}_{R}\nu_{eL})(\nu_{aL}^{T}Ce_{L})\langle |H_{0}|\rangle +...,
\label{Leff}
\end{equation}
where the $\psi$s represent lepton SU(2) doublets, the $H$ is a doublet Higgs field with vacuum expectation value $\langle |H_{0}|\rangle$, and a large scaling factor for $\langle |H_{0}|\rangle$ is understood. The epsilon symbols are the constant SU(2)$_{L}$ antisymmetric two-index tensors and the SU(2)$_{L}$ doublet component indices are indicated as k, n, l and m. The lepton flavor label $a$ on the left-handed neutrino doublet in Eq.(\ref{Leff}), which can be $\mu$ or $\tau$, will be taken to be $\mu$ for definiteness. The working effective Lagrangian, which includes the SM part and the $\Delta$L=2 NSI part, reads

\begin{equation}
{\cal L}_{eff}=-4\frac{G_{F}}{\surd 2}\left[(\bar{e}_{L}\gamma_{\lambda}\nu_{eL})(\bar{\nu}_{\mu L}\gamma^{\lambda}\mu_{L})+2\epsilon(\bar{\mu}_{R}\nu_{eL})(\nu_{\mu L}^{T}{C}e_{L})\right] 
\label{Leff2}
\end{equation}
The first term in Eq.(\ref{Leff2}) is the SM, charged current effective Lagrangian, while the second term is the NSI, lepton-number violating term. As remarked earlier, the factor 2 in the definition of $\epsilon$ compensates for the factor 1/4 in the decay rate with the S,P structure compared to the V-A structure. When $\epsilon$=1, the rates of the two types of interaction are the same with this definition. The parameter $\epsilon$ in Eq.(\ref{Leff2}) is given in terms of the underlying model's parameters as defined in Eq.(\ref{B13}) by the expression given in Eq.(\ref{ep}).  

\subsection{Fluxes}

The neutrino fluxes from DAR are isotropic, with energy distributions that are determined by the distributions of the individual muon decays.  Given the same coupling coefficient in the effective Lagrangian, the total rates of the Lorentz scalar decays are 1/4 those of the vector decays, but the shape of the energy distribution of the $\nu_{e}$ in the SM decays is the same as that in the NSI decay and the same is true of the distribution of $\nu_{\mu}$ in the decays.  Normalized to the LSND total flux averaged over the volume of their detector, the flux rates in $cm^{-2}MeV^{-1}$ units as a function of the neutrino energy are

\begin{equation}
\frac{d\phi(E)}{dE}=5.7\times10^{13}\times(\frac{E}{m_{\mu}})^2(1-\frac{4E}{3m_{\mu}})\Theta(1-\frac{2E}{m_{\mu}})
\label{eq:V-Aflux}
\end{equation}
for the $\bar{\nu}_{\mu}$ flux, and 
\begin{equation}
\frac{d\phi(E)}{dE}=5.7\times10^{13}\times(\frac{E}{m_{\mu}})^2(2-\frac{4E}{m_{\mu}})\Theta(1-\frac{2E}{m_{\mu}})
\label{eq:V+Aflux}
\end{equation}
for the $\bar{\nu}_{e}$ flux. The corresponding KARMEN fluxes are a factor 0.55 smaller than LSND, so the prefactor 5.7 in the flux expressions above is 3.1 in the KARMEN experiment.  When LSND energy resolution is included, these distributions extend up to 60 MeV, as will be clear later in the event rate plots. The KARMEN resolution is much sharper and has little effect on the flux distributions or event rate distributions.

\subsection{$\bar{\nu}_{\mu}$ and $\bar{\nu}_{e}$ Oscillation Probabilities}

As we stressed in the Introduction, both appearance and disappearance of $\bar{\nu}_{e}$ play a role in generating the inverse $\beta$-decay production of an $e^{+}$ signal in the detectors. We include the NSI parameter $\epsilon$ in the oscillation probabilities, though they could just as well be put in the flux factors.  Denoting the probability of appearance of $\bar{\nu}_{e}$ from oscillation of $\bar{\nu}_{\mu}$ as $P_{\mu e}$ and the survival of $\bar{\nu}_{e}$ after its direct production in $\mu$-decay as $P_{ee}$, we write
\begin{eqnarray}
 P_{\mu e}(E_{\nu\mu}) &=& 4|U_{e4}|^2|U_{\mu 4}|^2 \sin^2(\Delta(E_{\nu\mu}))(1+|\epsilon|^2) \nonumber \\
 P_{ee}(E_{\nu e})      &=& |\epsilon|^2( 1-4(|U_{e4}|^2(1-|U_{e4}|^2))\sin^2(\Delta(E_{\nu e}))),
 \label{oscillationPs}
 \end{eqnarray}
where we define $\Delta(E_{\nu i})=1.27 M_{4}^2[eV^{2}] L[m]/(4 E_{\nu i}[MeV])$, $i$ = e, $\mu$ or $\tau$, and $M_{4}^2$ stands for the dominant difference in the square of the sterile neutrino mass and that of a standard 3-neutrino mixing mass eigenstate value. The value of L is 30m $\pm$ 4m  (LSND) or 17m $\pm$ 1m (KARMEN).
In modeling the event rate, we have two contributions.  There is one originating from the $\bar{\nu}_{\mu}$ component of the flux, with a probability of $P_{\mu e}(E_{\nu\mu})$ of oscillating to $\bar{\nu}_{e}$ plus a contribution from the NSI generated $\bar{\nu}_{e}$ component of the flux that survives to reach the detector with a probability $P_{ee}(E_{\nu e})$.

In the absence of NSI, the $\bar{\nu}_e$ appearance is driven solely by the light, sterile-neutrino oscillation term.  In this case there are only two effective parameters, $4 |U_{e4}|^2 |U_{\mu 4}|^2 \equiv \sin^{2}(2\theta_{\mu e})$ and $M_{4}^{2}$. \emph{ Only the product of the two mixing parameters is determined by a fit to appearance data.} This will be important in our study of the simultaneous fits to the combined LSND and KARMEN data.

\section{Expected Event Rates}

\begin{figure}[htbp]
   \centering 
   \includegraphics[width=4in]{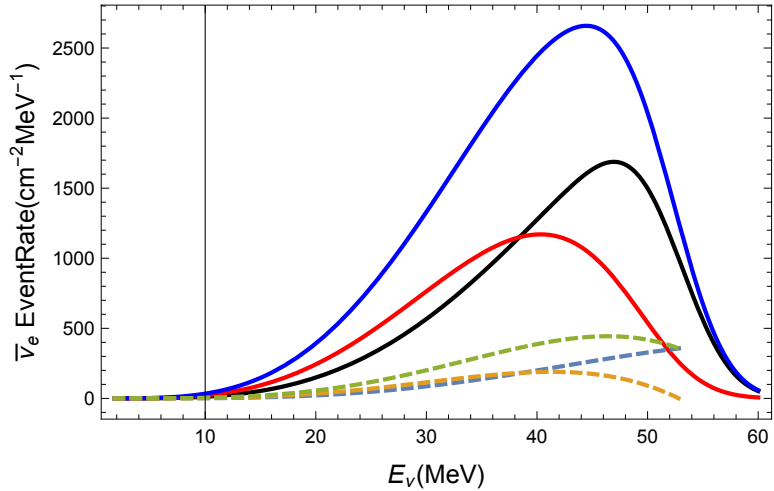} 
   \caption{The solid curves are LSND events: total (top, blue), $\bar{\nu}_{\mu}$ at $\mu$-decay source (middle, black) and $\bar{\nu}_{e}$ at $\mu$-decay source (bottom,red). The dashed curves are KARMEN events: total (top, green), $\bar{\nu}_{\mu}$ at $\mu$-decay source (middle, light blue) and and $\bar{\nu}_{e}$ at $\mu$-decay source (bottom, orange). }
   \label{fig:all_events_plot}
\end{figure}

The event rates expected depend upon the flux of incoming neutrinos, the number of proton targets, the inverse beta decay cross section, $\sigma_{IBD}(E_{\bar{\nu} e})$ \cite{vogelandbeacom} and the probabilities that the source neutrinos provide $\bar{\nu}_{e}$s after oscillation. Putting it all together, the energy distribution of events for the whole data set as a function of energy can be written

\begin{equation}
\frac{dN}{dE_{\bar{\nu}}}=\tilde{\varepsilon}\times N_{p}\times\sum_{i=e,\mu} \frac{d\Phi}{dE_{\nu i}}\times \sigma_{IBD}(E_{\bar{\nu} e})\times P_{i e}(E_{\bar{\nu} i}),
\label{eqndiffRate}
\end{equation}
where $\tilde{\varepsilon}$ is the average efficiency for detection, $N_{p}$ is the number of proton targets and the sum indicates that there are two sources of $\bar{\nu}_{e}$s to trigger the signal events. If all of the $\bar{\nu}_{\mu}$ were to oscillate to $\bar{\nu}_{e}$ and the probability that the $\bar{\nu}_{e}$ would be produced at the source and survive to reach the detector were 1, the distribution of events with respect to neutrino energy that would result is shown in Fig. \ref{fig:all_events_plot}. The LSND spectra are illustrated by the solid curves and the KARMEN spectra are illustrated by the dashed curves.  In each case, the top curve shows the sum of the two below it, where the middle (above 40 MeV) curve, shows the contribution of $\bar{\nu}_{\mu}$ from the source and the lower curve shows the contribution of $\bar{\nu}_{e}$ from the source. 

The curves show the hypothetical energy distribution of events that results when $P_{\mu e}=P_{ee}$ =  1 in Eq.(\ref{eqndiffRate}). 
To illustrate the meaning of the curves in Fig. \ref{fig:all_events_plot}, we note that the integral of the solid black curve (middle) from 20 MeV to 60 MeV represents the total possible number of observable events in the LSND experiment \cite{LSND1}, which they estimate as 33,300 $\pm$ 3300 events.  With 88 $\pm$ 22 (statistical) events observed above background, they report the ratios $P_{\mu e}$ = 0.00264 $\pm$ 0.00067. Our modeling of the distribution (the black curve in Fig. \ref{fig:all_events_plot}) of their value of 33,300 expected for 100\% transmutation of $\bar{\nu}_{\mu}$ to $\bar{\nu}_{e}$ is 34,300, well within their uncertainties.  The size of the expected sample from KARMEN is smaller than LSND's because their total flux, their number of target nuclei and their reconstruction efficiency are all smaller.

With pure oscillation, no NSI, the middle curves are the relevant ones when computing the total oscillation probability based on an observed signal of $\bar{\nu}_{e}+p \rightarrow e^{+}+n$ events at the detector. Note that the extension of the LSND curves up to 60 MeV results from their broad energy resolution.

\section{Numerical Results}

In this section we present $\chi^{2}$ fits to the LSND and KARMEN data in our models with four parameters $|U_{e4}|, |U_{\mu 4}|, M_{4}^{2}$ and $\epsilon$. For LSND \cite{LSND2} we consider the total number of events above background and the cleaner subsample of the data used in Fig. 24 and  characterized in Table X of Ref.\cite{LSND2}. This subsample contains 32.2 $\pm$ 9.4 events with\footnote{The measure $R_{\gamma}$ is defined in Sec. VII C of Ref. \cite{LSND2} as the likelihood that the $\gamma$ is correlated with the prompt gammas from positron annihilation divided by the likelihood that it is accidental.} $R_{\gamma} > 10$.  For KARMEN we use the distribution of events vs. energy in Fig. 11b of Ref.\cite{KARMEN}. The LSND oscillations subsample and the KARMEN data that we use in our combined fits are shown in Fig. 4.

\subsection{Unconstrained fit}
\begin{figure}[htbp]
\centering
\includegraphics[width=6in]{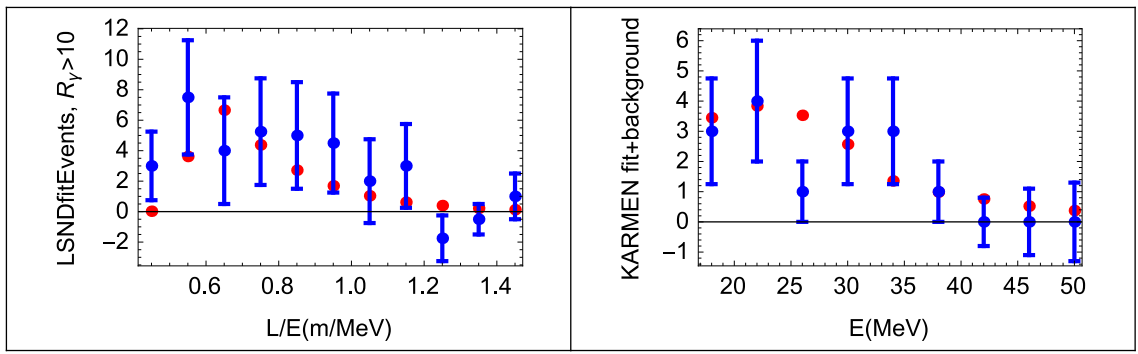}
\caption{Left panel: The LSND events above background, with error bars, are shown in blue in bins of width 0.1 m/MeV. The corresponding best fit expectations for the bins are shown as red dots. There is no normalization constraint.
Right panel: The KARMEN events, with error bars, are shown in blue in bins of width 4 MeV. The corresponding best fit to background plus model contribution for the bins is shown as red dots.}
\label{fig:jointplot}
\end{figure}

Minimizing a joint LSND and KARMEN $\chi^{2}$ based on the 11 data bins in LSND's "excess events vs L/E" plot, Fig. 24 in Ref.\cite{LSND2} and the 9 data bins in KARMEN's "events/4 MeV vs. energy prompt event", Fig. 11b in Ref.\cite{KARMEN}, we find a "best fit" shown in Fig. \ref{fig:jointplot}. This fit's parameter values, $\chi^{2}$ per degree of freedom (d.o.f.) and the expected number of events given by the unconstrained fit are summarized in Table 2\footnote{We should mention that eliminating the $\bar{\nu}_{\mu}$ oscillations by setting $|U_{\mu 4}|$=0 and seeking the minimum of $\chi^{2}$ yields a solution at $\chi^{2}$=17.2, with $|U_{e4}|$=0.707,$M_{4}^2$ =11.2 $eV^{2}$ and $\epsilon$=0.025. Though the minimum is nearly degenerate with that of the unconstrained minimum, the mixing parameter values are far outside the bounds set by many other analyses \cite{glll2,glll1,whitepaper}.}.

\begin{table}[ht] 
\begin{center}
     \caption{Unconstrained best fit parameter values, $\chi^{2}$ per d.o.f. and fit value of excess LSND events. The best fit $\chi^{2}$ = 17.0 value is quite democratically split between KARMEN, 8.6, and LSND, 8.4. } 
     \smallskip
   \begin{tabular}{||l|l|l|l|l|l|lr||} 
 \hline 

      $|U_{e4}|$  &  $|U_{\mu 4}|$  & $M_{4}^{2}(eV^{2})$ & $\epsilon$ & $\chi^{2}/d.o.f.$ & excess LSND events\\  \hline
      
           0.144 &  0.101 &  4.64 & 0.0 & 17.0/(20-4) = 1.06 & 21.5 (32 in data)\\
 \hline               
       \end{tabular}
      \end{center}
\label{tab:freefit}
\end{table}

 The degeneracy we pointed out above in Sec. 2.2 comes into play here.  With a best fit that yields $\epsilon$ = 0, only the value of $|U_{e4}| \times |U_{\mu 4}|$ is determined by the (local) minimization procedure. The results shown for the mixing parameters are quite compatible with those reported in \cite{glll1} and \cite{glll2}, though the result we find for $M_{4}^{2}$ is high, a point we return to in the following section. The specific values of the mixing parameters are somewhat dependent on the choice of starting values chosen in the search for the minimum of  $\chi^{2}$. Any pair of values that yields $|U_{e4}| \times |U_{\mu 4}|$ = 0.0145 will find $|\chi^{2}|$= 17.0 with the accompanying values $M_{4}^{2}$=4.64 $eV^{2}$ and $\epsilon$=0.  Of course the amount of $\chi^{2}$ deviation from the best fit value that develops as one allows excursions in $M_{4}^{2}$ and/or $\epsilon$ with $|U_{e4}|$ and $|U_{\mu 4}|$ fixed, at their best fit values does depend on what pair of best fit values one chooses (consistent with the product value 0.0145). We explore excursions of parameters from best fit values next.\footnote{With the best fit parameters shown in Table 2, we note that the values of $|U_{e4}|$ and $M_{4}$ and their 68\% and 90\% C.L. extensions that we explore later allow for the possibility that the effective mass in neutrinoless double-$\beta$ decay may vanish \cite{dbeta1, dbeta2}.}

\subsection{$\Delta\chi^{2}$ deviations from best fits } 
\begin{figure}[htbp] 
   \centering 
   \includegraphics[width=3in]{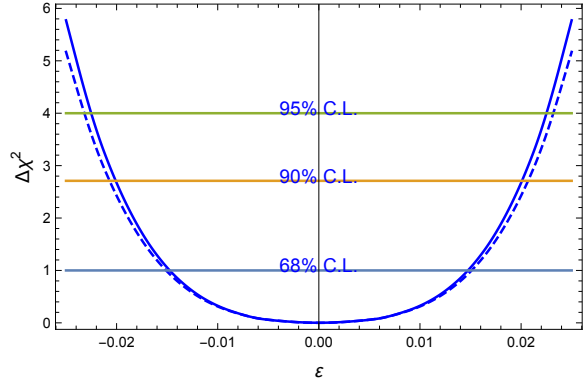} 
   \caption{The solid curve shows the deviation $\Delta \chi^2$ from best fit as a function of $\epsilon$ with other parameters fixed at their values shown in Table 2. Taking the equivalent best fit values $|U_{e4}|$=0.20 and $|U_{\mu 4}|$ = 0.0725, we find the deviation shown by the dashed curve. The horizontal lines indicate the 1$\sigma$, 90\%C.L. and 2$\sigma$ values for 1 degree of freedom.}
   \label{fig:delchi1}
\end{figure}

There are 32 LSND events represented in the left panel of Fig. \ref{fig:jointplot}, while the fit represents 21.5 events. Even with an uncertainty of 9 in the number of events, without a constraint on the normalization of the fit  there is evidently tension between L/E dependence of the data, the oscillation L/E modeling of the data and the overall normalization of the fit. Pursuing this question, we keep the values of $|U_{e4}|$, $|U_{\mu 4}|$ and $M_{4}^{2}$ fixed at the best fit values shown in Table 2 and determine the effect on $\chi^{2}$ of letting $\epsilon$, our lepton number violation parameter, deviate from it's best fit value of 0. The result is shown in Fig. \ref{fig:delchi1}.   At $\epsilon$ = 0.015, $\Delta \chi^{2}$ = 1, 68\% C.L., and at $\epsilon$ = 0.02, $\Delta \chi^{2}$ = 2.71 , 90\% C.L.. The corresponding values of the number of expected events are 27 and 31, well within the experimental value 32 $\pm$ 9.  The dashed curve shows the result of choosing best fit values that have a larger $|U_{e4}|$ and correspondingly smaller $|U_{\mu 4}|$. The result shows very little sensitivity to individual values of the mixing parameters $|U_{e4}|$ and $|U_{\mu 4}|$.

\begin{figure}[htbp] 
   \centering 
   \includegraphics[width=6in]{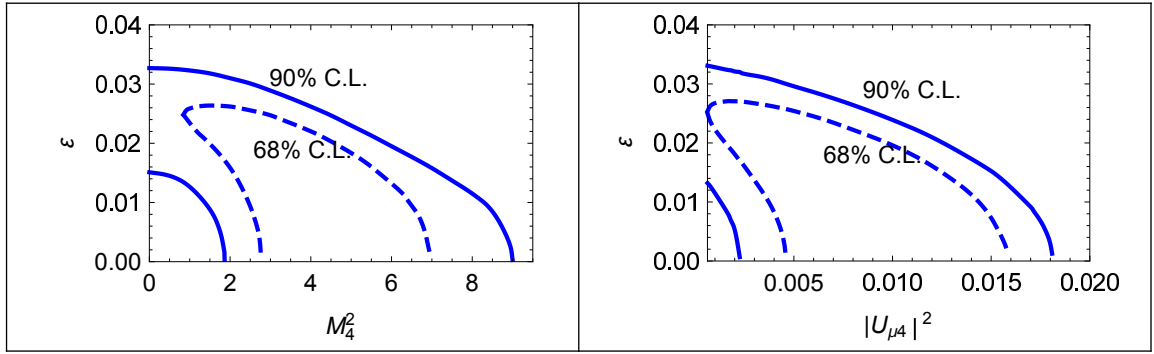} 
   \caption{The left-hand panel shows the 68\% and 90\% C.L., $\Delta \chi^2$ = 2.3 and 4.61 contours in the $\epsilon$ - $M_{4}^{2}(eV^{2})$ plane with mixing parameters chosen at best fit values $|U_{e4}|$ = 0.144 and $|U_{\mu 4}|$ = 0.101. The right-hand panel shows the 68\% and 90\% C.L., $\Delta \chi^2$ = 2.3 and 4.61, contours in the $\epsilon$ - $|U_{\mu 4}|^{2}$ plane with mixing parameters chosen at best fit values $|U_{e4}|$ = 0.144 and $M_{4}^{2}$=4.64 $eV^{2}$.}
      \label{fig:delchi2}
\end{figure}

\begin{figure}[htbp]    \centering 
   \includegraphics[width=4in]{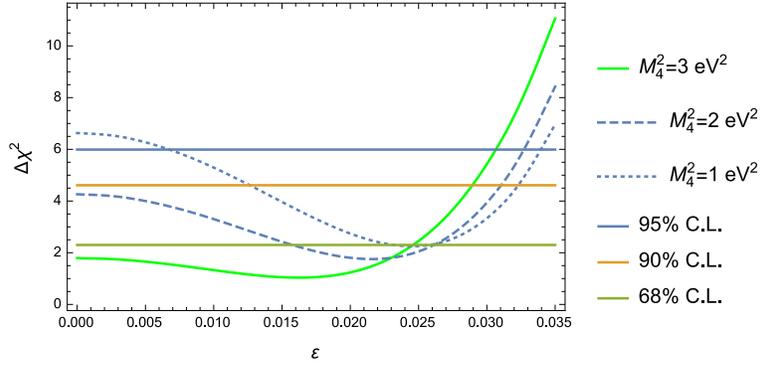} 
   \caption{The $\Delta \chi^2$ vs. $\epsilon$ projections of the left panel of Fig. \ref{fig:delchi2} at $M_{4}^{2}$ = 1, 2 and 3 $eV^{2}$, where mixing parameters are chosen at best fit values $|U_{e4}|$ = 0.144 and $|U_{\mu 4}|$ = 0.101. The constant $\Delta \chi^{2}$ lines are at 2.3, 4.61 and 5.99, corresponding to 68\%, 90\% and 95\% C.L. boundaries for 2 degrees of freedom. Choosing other values for $|U_{e4}|$ and $|U_{\mu 4}|$ consistent with the product 0.0145 gives essentially the same results.}
   \label{fig:delchi3}
\end{figure}

Pursuing this point further, we first allow both $\epsilon$ and $M_{4}^{2}$ to take on a range of values away from those of the best fit to see if reasonable fits with smaller $M_{4}^{2}$ values can be found that are more compatible with the global fit, which is in the 1.5 - 2 e$V^{2}$ range \cite{glll2}.  We fix the best fit values of the mixing parameters $|U_{\mu 4}|$ and $|U_{e4}|$ at the values given in Table 2, and let $M_{4}^{2}$ and $\epsilon$ vary to find the 68\%, 90\% etc. allowed regions in their 2-degrees of freedom space. The result is shown in the left-hand panel of Fig. \ref{fig:delchi2}.   Again, to check the sensitivity of the results to the choice of mixing parameters, we set $|U_{\mu 4}|$ equal to 0.0145/$|U_{e4}|$ to satisfy the minimization condition.  With $|U_{e4}|$ = 0.20 and $|U_{\mu 4}|$ = 0.0725, for example, we find that the 68\% C.L. and 90\% C.L. contours are within the width of the contour lines shown in the left panel of Fig. \ref{fig:delchi2}.   Adding detail to the left-hand plot of Fig. \ref{fig:delchi2}, we plot $\Delta \chi^{2}$ vs. $\epsilon$ profiles at constant $M_{4}^{2}$ for 1, 2 and 3 $eV^2$ in Fig. \ref{fig:delchi3}.  Both plots show that for $M_{4}^{2}$ above 1 $eV^2$,  there is substantial parameter space consistent with global fits where the $\Delta$L-breaking effects plus sterile neutrino oscillations accommodate the LSND and KARMEN data, even at 1$\sigma$.      

Similarly, the right-hand plot in Fig. \ref{fig:delchi2} shows that values compatible with an interesting region of $|U_{\mu 4}|$ - $\epsilon$ values opens up within the 68\% and 90\% C.L. boundaries.  In particular, values of $\epsilon$ in the range $0.02 \leq \epsilon \leq 0.03$ indicated in Fig.\ref{fig:delchi2}, based on the LSND $R_{\gamma}$ data, agree with the allowed values we found in our study of the $\Delta$L $\neq$ 0 effects when comparing the SM predictions and precision measurements of Michel-parameters $\rho$, $\delta$ and $\xi$ in Section 2.3.   

\subsubsection{Comments on fits and on constraints on the fits}

Fitting the LSND data set alone, one finds that unconstrained, the 4-parameter fit produces $\epsilon$ = 0.  As in the combined fit, the value of the sterile neutrino mass scale are typically about 4 - 5 $eV^2$, larger by factors of 3 - 4 compared to the values of about 1-2$eV^{2}$ found in global fits \cite{glll2, kopp}. The LSND total and $R_{\gamma} \geq$ 10 excess events are somewhat poorly represented by the best fit parameter set.  As in the joint fit, the individual fits have $\chi^{2}$ of 8 to 9, $\sim$1 per d.o.f. If one demands a highly constrained fit, one that stays close to that of a global, 1- sterile neutrino model fit, the "HIG" set of Table I in Ref. \cite{glll2} for example, and requires a value of $\epsilon$ that produces LSND's central value of 88 events, namely $\epsilon$ = 0.056, the resulting $\chi^{2}$ for the L/E distribution, Fig. 24 in Ref. \cite{LSND2}, is 10.7 for 11 d.o.f., which is about the same as the LSND contribution to the best fit to the joint distribution. This constrained parameter set gives a very poor description of the KARMEN data, as one expects.  However, if one allows a 90\% C.L. excursion from the 88 $\pm$ 22 excess LSND events, one finds the value $\epsilon$ = 0.041, which is consistent with the 90\% C.L. bound of signal events found with KARMEN's 15 events with a background of 15.8. (Since KARMEN observed no events in the 3 highest energy bins, at 42 MeV, 46 MeV and 50 MeV, it reports zero $\surd N$ errors in those bins. To estimate a statistically expected range for these points, we adopt the Poisson 68\% C.L. signal mean for no events observed with the KARMEN background numbers given for those bins \cite{fandc}.)

\begin{figure}[htbp] 
   \centering 
   \includegraphics[width=4in]{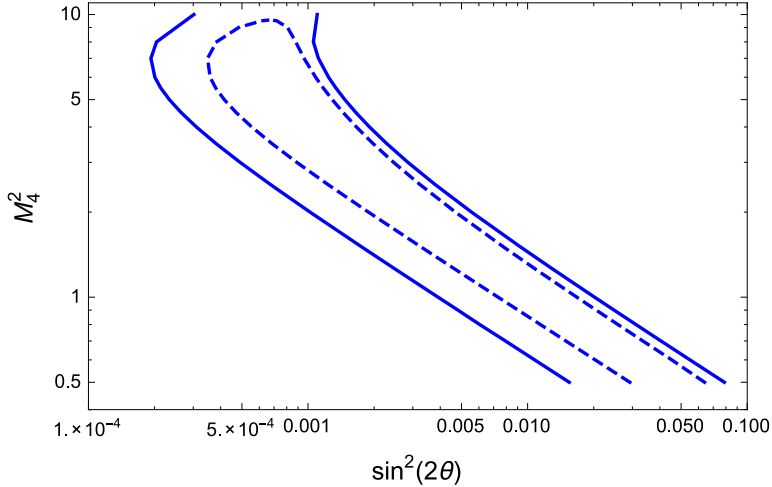} 
   \caption{Our boundaries at 68\%C.L. and 90\% C.L in $\sin^{2}(2\theta_{\mu e})$ - $M_{4}^{2}$ space with $\epsilon$ = 0, its best fit point value, where $\sin^{2}(2\theta_{\mu e})$ = 0.00084 and $M_{4}^{2}$ = 5.1$eV^{2}$.}
   \label{fig:MsqVSssq}
\end{figure}

Comparing our results directly with the DAR results for LSND as presented in the highly detailed statistical analysis of Ref.\cite{cems02}, we do not expect close agreement because their data sample is much larger, cut at $R_{\gamma} \geq 10^{-5}$ rather than at 10, and the analysis in \cite{cems02} works directly with the data, while we use a subsample that has been refined and binned and presented graphically in the final LSND paper \cite{LSND2}.  Our version of their data, Fig. \ref{fig:jointplot}, is obtained from reading off the bins and error bars in Fig. 24 of Ref.\cite{LSND2} and checking that the number of events and the corresponding statistical uncertainty agrees with the $R_{\gamma} \geq$10 values listed there in Table X.  Nonetheless, it is useful to compare Fig. 6 in Ref.\cite{cems02} with a similar version generated from our fit.  In Fig. \ref{fig:MsqVSssq}, we show the 68\% and 90\% C.L contours in the variables $M_{4}^{2}$ vs. $\sin^{2}(2\theta_{\mu e})$ with $\epsilon$ = 0, it's value at best fit point where the $M_{4}^{2}$ and $\sin^{2}(2\theta_{\mu e})$ values are 5.1 $eV^{2}$ and 0.00084. The region of overlap lies below $M_{4}^{2}$ = 2 $eV^{2}$, where our 90\% C.L. region covers their 95\% region. At higher values of $M_{4}^{2}$, our 90\% region lies at smaller $\sin^{2}(2\theta_{\mu e})$ values than theirs.  Our rough analysis, designed to explore a combined $\Delta$L $\neq$ 0 and sterile neutrino mixing modeling of DAR appearance data does not capture the high-mass region details presented in \cite{cems02}, but it does overlap with the region where $0.02 \leq \epsilon \leq 0.03$, is allowed by LSND, KARMEN and precision $\mu$-decay data. 

\subsubsection{Comments on Future Possibilities}

Looking ahead to (far) future possibilities where leptonic, $\Delta$L = 2 interactions may reveal themselves, we suggest that a neutrino factory would be ideal.  The short baseline scheme proposed by Giunti, Laveder and Winter \cite{glw}, for example, emphasizes the clean environment for studying new physics in both charge modes of the muon at a $\nu$-factory. Their focus is on oscillation studies, but the short baselines proposed, 50 m and 200 m, lend themselves to study of excess events originating from both of the exotic decays $\mu^{-} \rightarrow e^{-}+\nu_{e}+\nu_{\mu}$ and $\mu^{+} \rightarrow e^{+}+\bar{\nu_{e}}+\bar{\nu}_{\mu}$, where both $\Delta$L=2 and oscillation effects can also be looked for. 

In the shorter term, planned high precision DAR experiments that include $\mu^{+}$ decays at rest would provide welcome tests for our scheme.  The JSNS$^{2}$ experiment \cite{jsns2,FS} with a 24m baseline setup similar to LSND plans to look for $\bar{\nu}_{e}$ appearance from $\pi^{+}$, K$^{+}$ and $\mu^{+}$ decay at rest. All three are sensitive to a sterile neutrino effects, but only the $\mu^{+}$ decay provides a test of our $\Delta$L = 2 NSI. Therefore any difference from sterile-only effects that shows up in the $\mu^{+}$ case as compared to the hadronic decay sources would be evidence for purely leptonic new physics. Likewise, to the extent that $\mu$ decays contributing to the atmospheric neutrino signals \cite{IceCube} can be distinguished from $\pi$ and K decays, there is a possibility that predictions of our model can be tested. 

In addition, within specific renormalizable models that realize the anomalous $\mu$-decay via $\Delta$L$\neq$ 0 four-Fermion interactions, there are striking signatures that would support the whole picture.  Examples were given in Sec.(2.4) of effects on the $e^{+}e^{-}\rightarrow\mu^{+}\mu^{-}$ cross section and of direct production and decay of new neutral and charged scalars at the LHC.

\section{Summary and conclusions}
To apply the direct $\Delta$L = 2, lepton number violating interaction Eq. (\ref{Leff2}) in $\mu$-decay toward explaining LSND \cite{LSND2} anomaly \cite{bands} without contradicting the absence of oscillation signal in KARMEN data \cite{KARMEN}, we added mixing between the known neutrino flavors and a sterile neutrino \cite{ggms}. Unlike the new interaction, this affects the signals in the two experiments differently because of their difference in baselines.  We first reviewed the lepton \emph{flavor} violating, but lepton number conserving cases that, when generalized to SU(2) symmetric models,  are too strictly constrained by limits on charged lepton flavor violation to admit the LSND signal.  We then constructed a set of nine candidate $\Delta$L $\neq$ 0, consistent effective interactions that are not constrained by charged lepton data.  Seven of the nine were shown to be disfavored by the precision measurements of the $\mu$-decay Michel parameters.  Two cases make no change to the SM tree-level values $\rho$=$\delta$=3/4, consistent with the current data and produce better overall fits with an  $\epsilon$ parameter value that is consistent with the range allowed by our fitting of the combined LSND and KARMEN DAR data sets represented by the binned distributions in Fig. 6.\footnote{Apart from Ref.\cite{cems02}, all fits that we're aware of contain data beyond the DAR data, which are the \emph{only} data sensitive to our new physics, $\Delta$L $\neq$ 0 interactions.} We then constructed renormalizable models for both of the Michel parameter preferred effective Lagrangians, connecting the underlying coupling constants and scalar masses to the $\epsilon$ parameter.  One of these  was chosen for a detailed phenomenological study. The new physics interaction, whose strength is parameterized by $\epsilon$, produces the $\Delta$L= 2 decay mode $\mu^{+} \rightarrow e^{+} + \bar{\nu}_{\mu} + \bar{\nu}_{e}$. This example is compatible with experimental limits on deviations from the standard model description of $\mu$ decay \cite{kando} and has the special feature that there are two sources of $\bar{\nu}_{e}$, those from the appearance oscillation of $\bar{\nu}_{\mu}$ to $\bar{\nu}_{e}$ and the survival of the NSI-generated $\bar{\nu}_{e}$. We studied the interplay between the two contributions and the relative roles that the sterile-neutrino-driven oscillations and the $\Delta$L=2 new physics played. 

We found a good, unconstrained fit to the combined DAR data sets with a fourth, sterile neutrino oscillating with parameter set $|U_{e4}|$ = 0.144, $|U_{\mu 4}|$ = 0.101, and $M_{4}^2$ = 4.64 $eV^{2}$ and the NSI parameter $\epsilon$ = 0, see Table 2. The mixing parameter values are quite compatible with fits to appearance data \cite{glll1} and global fits to both appearance and disappearance data \cite{glll2}, and, though the mass scale is higher than best fits reported for the combined, global fits \cite{glll2}, we show our 90\%C.L. contour in Fig. \ref{fig:MsqVSssq}, which overlaps that of \cite{cems02} for masses below 2 $eV^2$ and is thus consistent with global fits. Calculating the number of excess events with our best fit parameters produces an estimate of 21, whereas the data subsample contains 32 $\pm$ 9 events.  An analysis of the $\Delta \chi^{2}$ vs. $\epsilon$ shows that for $2\; eV^{2} \geq M_{4}^{2} \geq 1\; eV^{2}$ there are points in the ranges $0.015 \leq \epsilon \leq 0.026$ allowed at the 68\% C.L and points in the range $0 \leq \epsilon \leq 0.032$  allowed at 90\% C.L.. Many of the points in these regions bring the modeled numbers of events for LSND into much better alignment with the data, while staying within the constraints from KARMEN.  A similar improvement is found with $R_{\gamma} \geq 1$ and the total number of excess events is computed for LSND so the fit to the shapes of the event distributions favors a simple 1-sterile neutrino picture, but the numbers of events in the LSND experiment are marginally reproduced and an added contribution from the $\Delta$L = 2 interaction within the 68\% to 90\% C.L. limits greatly improves the overall rates.  

To recap and conclude, we find that values of the sterile neutrino mixing parameters are reasonably close to those from a global fit reported recently in Ref. \cite{glll2}, and the mass-squared splitting  $M_{4}{^2}$ found in our fit, though larger than the global-fit splitting parameter, is compatible at 68\% to 90\% C.L with the 1-2 $eV^2$ region of the detailed LSND fits.  Moreover, our constrained fit investigations indicate that LSND fits with non-zero $\epsilon$ values and smaller values of $M_{4}^{2}$ give acceptable fits, suggesting that the $\Delta$L = 2 NSI may still be playing a role with values up to 0.025-0.03, in accord with the constraints imposed by Michel parameter measurements. We conclude that this class of NSIs brings in new features of the sterile neutrino idea and it is well worth pursuing in analyzing current and future neutrino data and, at the underlying model level, current and future collider data.

\bigskip
\emph{Acknowledgments:} We thank KITP Santa Barbara for hospitality and support during the "Present and Future Neutrino Physics" workshop, where this work was started. D. McKay and K. Babu thank the lively CETUP 2015 workshop for providing the opportunity to present and discuss this work, and especially Barbara Szczerbinska, who kept it all together.  We appreciate helpful communications with Bill Louis, Geoff Mills and Thomas Schwetz. This work is supported in part by the U.S. Department of Energy Grant No. DE-SC0010108 (KSB)  and by DE-SC0013699 (IM).

\section{Appendix}
In this appendix we collect the formulas and definitions from Ref. \cite{kando} that we use in the developments of Sec.2.3.
The basic idea is simply that models that produce a final state positron spectrum with the same shape as that of the V-A form of the SM (see Eq.(\ref{eq:V-Aflux})) will be compatible with the high-precision data, which is compatible with the SM values of Michel parameters \cite{PDG14, twist12}.  Those models that give the other $e^{+}$ spectral behavior (see Eq.(\ref{eq:V+Aflux})) will add terms that disagree with the prediction of the V-A interaction of the SM. With appropriate modification of the leptonic labels, the Michel parameters  for any model can be expressed in terms of the coupling coefficients of the following generic 4-Fermi effective Lagrangian:
\begin{eqnarray}
{\cal L}_{\mu\rightarrow e\nu\nu}&=&-\frac{4G_{F}}{\surd2}[g_{RR}^{S}(\bar{e}_{R}\nu_{eL})(\bar{\nu}_{\mu L}\mu_{R})+R\leftrightarrow L+g_{RL}^{S}(\bar{e}_{R}\nu_{eL})(\bar{\nu}_{\mu R}\mu_{L})+R\leftrightarrow L \nonumber \\ 
                                           & &\mbox{}+ g_{RR}^{V}(\bar{e}_{R}\gamma^{\mu}\nu_{eR})(\bar{\nu}_{eR}\gamma_{\mu}\mu_{R})+R\rightarrow L+g_{RL}^{V}(\bar{e}_{R}\gamma^{\mu}\nu_{eR})(\bar{\nu}_{\mu L}\gamma_{\mu}\mu_{L})+R\leftrightarrow L + \nonumber \\                     
                                            & &\mbox{}+\frac{g_{RL}^{T}}{2}(\bar{e}_{R}\sigma^{\mu\nu}\nu_{eL})(\bar{\nu}_{\mu R}\sigma_{\mu\nu}\mu_{L})+R\leftrightarrow L]+H.c..
\label{eq:genL}
\end{eqnarray}

Calculation of the decay rate imposes the normalization condition

\begin{eqnarray}
1&=&\frac{1}{4}(|g_{RR}^{S}|^{2}+|g_{LL}^S|^{2}+|g_{RL}^S|^{2}+|g_{LR}^{S}|^{2})+(|g_{RR}^{V}|^{2}+|g_{LL}^{V}|^{2} \nonumber \\ 
  & &\mbox{}+|g_{RL}^{V}|^{2}+|g_{LR}^{V}|^{2})+3(|g_{RL}^T|^{2}+|g_{LR}^T|^{2}).
  \label{eq:norm}
  \end{eqnarray}
  
  Using Eq( \ref{eq:norm}) to express $|g_{LL}^{V}|^{2}$ in terms of the other, presumed small, coupling constants, substituting it into the appropriate equations for the Michel parameters as summarized in the review "Muon Decay Parameters" in Ref.\cite{PDG14} and then keeping leading order terms in modulus squared couplings, one finds the following 
expressions for the $\rho$, $\delta$ and $\xi$-parameters \cite{kando}:
\begin{equation}
\rho=\frac{3}{4}\left[1-\left(|g_{RL}^{V}|^{2}+|g_{LR}^{V}|^{2}+2|g_{RL}^{T}|^{2}+2|g_{LR}^{T}|^{2}+Re(g_{RL}^{S}g_{RL}^{T*}
+g_{LR}^{S}g_{LR}^{T*})\right)\right].
\label{eq:rho}
\end{equation}

\begin{equation}
\delta=\frac{3}{4}\left[1+3\left(|g_{RL}^{V}|^{2}-|g_{LR}^{V}|^{2}+2(|g_{RL}^{T}|^{2}-|g_{LR}^{T}|^{2})+Re(g_{RL}^{S}g_{RL}^{T*}
-g_{LR}^{S}g_{LR}^{T*}\right)\right)].
\label{eq:delta}
\end{equation}

\begin{eqnarray}
\xi&=&1+(-\frac{1}{2}|g_{RR}^{S}|^2-\frac{1}{2}|g_{LR}^S|^{2}-2|g_{RR}^{V}|^{2}-4|g_{RL}^V|^{2}+2|g_{LR}^{V}|^{2}+2|g_{LR}^{T}|^{2}  \nonumber \\
    & &\mbox{} -8|g_{RL}^{T}|^{2}-4Re(g_{RL}^{S}g_{RL}^{T*}-g_{LR}^{S}g_{LR}^{T*})).
\label{eq:xi}
\end{eqnarray}

The model $(B_{1})_{3}$, where the first term in Eq.(10) with a=$\mu$ or $\tau$ is the relevant one for our study, corresponds to the term with coupling $g_{LR}^{S}$ in Eq.(\ref{eq:genL}) and in Eq.(\ref{eq:rho}), where this coupling appears only in the term multiplied by $g_{LR}^{T}$. The latter coupling is not present in this model, so the $\rho$ and $\delta$ parameters retain the SM value of 3/4.

To illustrate the role that possible SU(2)$_{L}$ breaking might play, we introduce a symmetry breaking parameter c by rewriting Eq.(8) as $(L_{B.1})_{2}\rightarrow (\bar{\mu}_{R}\nu_{eL})(\nu_{aL}^{T}Ce_{L})-c\times(\bar{\mu}_{R}e_{L})(\nu_{aL}^{T}C\nu_{eL})$, where c is complex in general.  After a Fierz reordering to match the convention of Eq.(\ref{eq:genL}) and collecting terms, we find

\begin{equation}
({\cal L}_{B.1})_{2}\rightarrow (1+\frac{c}{2})(\bar{\mu}_{R}\nu_{eL})(\nu_{aL}^{T}Ce_{L})+\frac{c}{8}(\bar{\mu}_{R}\sigma_{\mu\nu}\nu_{eL})(\nu_{aL}^{T}{C}\sigma^{\mu\nu}e_{L}).
\label{eq:symbrk}
\end{equation}
Making the identifications $g_{LR}^{S}=N(1+\frac{c*}{2})\epsilon^{*}$ and $\frac{g_{LR}^{T}}{2}=N\frac{c^{*}\epsilon^{*}}{8}$, we choose the normalization factor N so that it insures the $\epsilon\rightarrow$ 0 limit leaves only $({\cal L}_{B.1})_{2}$ to describe $\mu$-decay . This requires that $N=\frac{2}{\surd{1+|\epsilon|^2+Re(c)}}$, and using Eq.(\ref{eq:rho}), we find $\rho=\frac{3}{4}(1-|\epsilon|^{2}\frac{|c|^2+Re \;c}{\surd{1+|\epsilon|^2+Re(c)}})$. For c = 1, there is no symmetry breaking and the coefficient of $|\epsilon|^{2}$ is $\frac{2}{3}$, as listed in Table (1).


\begin{thebibliography}{99}

\bibitem{LSND1} C. Athanassopoulos {\it et al.}, Phys. Rev. Lett. {\bf 75}, 2650 (1995); C. Athanassopoulos {it et al.}, Phys. Rev. Lett. {\bf 81}, 1774 (1998).

\bibitem{solarandatmo} J. N. Bahcall, P. I. Krastev, and A. Yu. Smirnov, Phys. Rev. D, {\bf 58}, 096016 (1998); S. M. Bilenky and C.Giunti, Phys. Lett B {\bf 444}, 379 (1998).

\bibitem{jm} L.M. Johnson and D.W. McKay, Phys. Lett B {\bf 433}, 355 (1998). 

\bibitem{bands} K.S. Babu and Sandip Pakvasa, "Lepton Number Violating Muon Decay and the LSND Anomaly", arXiv: hep-ph/020423v1 (2002).

\bibitem{barger} V. Barger {\it et al.}, Phys. Lett. B {\bf 489}, 345 (2000).

\bibitem{conrad} M.~Sorel, J.~M.~Conrad, and M.~Shaevitz, Phys. Rev. D {\bf 70}, 073004 (2004); A.~J.~Anderson, J.~M.~Conrad, E.~Figuaro-Feliciano, C.~Ignarra, G.~Karagiorgi,  K.~Scholberg, M.~H.~Shaevitz and J.~Spitz, Phys. Rev. D {\bf 86}, 013004 (2012).

 \bibitem{dlp} H. Davoudiasl, P. Langacker and M. Perelstein, Phys. Rev. D {\bf 65}, 105015 (2002) .
 
 \bibitem{bandm} G. Barenboim and N.E. Mavromatos, JHEP {\bf 0501}, 034 (2005).

\bibitem{kopp} J.~Kopp, P.~A.~N.~Machado, M.~Maltoni and T.~Schwetz, JHEP {\bf 1305}, 050 (2013).

\bibitem{glll2} C. Giunti, M. Laveder, Y.F.Li and W.W. Long, Phys. Rev. D. {\bf 88}, 073008 (2013).

\bibitem{ggms} M.~C.~Gonzalez-Garcia, M.~Maltoni and T.~Schwetz, arXiv:1512.06856 [hep-ph].  This reference provides an up-to-date review of the sterile neutrino application to short baseline anomalies.

\bibitem{gariazzo} S.~Gariazzo, C.~Giunti, M.~Laveder, Y.~F.~Li and E.~M.~Zavanin, J. Phys. G {\bf 43}, 033001 (2016).
 
 \bibitem{glll1} C. Giunti, M. Laveder, Y.~F.~Lie and W.~W.~Long, Phys. Rev. D {\bf 86}, 113014 (2012).
 
 \bibitem{SK15} K. Abe {\it et al.} (The Super Kamiokande Collaboration), Phys. Rev. D {\bf 91}, 052019 (2015) provides a bound $|U_{\mu4}| \leq 0.18$ for $M_{4}^{2} \geq 0.1 eV^{2}$, consistent with the fits of \cite{glll2} and the ranges of interest to us.

\bibitem{T2K15} K. Abe {\it et al.} (The T2K Collaboration), Phys. Rev. D {\bf 91}, 051102 (2015) uses its near detector to study $\nu_{e}$ disappearance. The bound $\sin^{2}(2\theta_{ee}) \leq 0.3$ for $\Delta M^{2} \geq 7 eV^{2}$ is found, equivalent to $|U_{e4}|^2 \leq 0.27$, again consistent with global fits and those we find.

 \bibitem{IC} Recent limits on $|U_{e4}|$ (see Eq. \ref{oscillationPs}) from analysis of IceCube data are quite stringent: "Sterile Neutrinos in Cold Climates", Benjamin J. P. Jones (MIT), FERMILAB-THESIS-2-15-17 (Experiment:FNAL-0974); M. Lindner, W. Rodejohann and X.-J. Xu, JHEP {\bf 1601}, 124 (2016). Because both analyses make reference to the MiniBoone results as well as to LSND, it is not clear how they relate to our fits, which refer only to LSND and KARMEN DAR data. Nominally, our fit gives $\sin^2(2\theta)$ = 0.00084 and $M_{4}^{2}$ = 4.6 $eV^2$.
 
 \bibitem{whitepaper} "Light Sterile Neutrinos: a Whitepaper", K.N. Abazajian {\it et al.}, arXiv:1204.5379v1 [hep-ph] (2012).
 
 \bibitem{bggns} J.~Bergstrom, M.~C.~Gonzalez-Garcia, V.~Niro  and J.~Salvado, JHEP {\bf 10}, 104 (2014).
 
\bibitem{noconcord} B. Leistedt, H. V. Peiris and L.Verde, Phys. Rev. Lett. {\bf 113}, 041301 (2014).

\bibitem{bandg99}  S. Bergmann and Y. Grossman, Phys. Rev. D {\bf 59}, 093005 (1999).

\bibitem{aands} E. Akhmedov and T. Schwetz, JHEP {\bf 1010}, 115 (2010).

\bibitem{twist11} R. Bayes {\it et al.} (The TWIST Collaboration), Phys. Rev. Lett. {\bf 106}, 041804 (2011).

\bibitem{twist12} A. Hillairest {\it et al.} (The TWIST Collaboration), Phys. Rev. D {\bf 85}, 092013 (2012).

\bibitem{mb} A. Aguilar-Arevalo {\it et al.}, (MiniBoone Collaboration), Phys. Rev. Lett. {\bf 110}, 161801 (2013).

\bibitem{LSND2} A. Aguilar {\it et al.} (LSND Collaboration), Phys. Rev. D {\bf 64}, 112007 (2001).

\bibitem{KARMEN} B. Armbruster {\it et al.} (KARMEN Collaboration), Phys. Rev. D {\bf 65}, 112001 (2002).

\bibitem{fgj} W. Fetscher, H. J. Gerber and K.F Johnson, Phys. Lett. B {\bf 173}, 102 (1986). 

\bibitem{kando} Y. Kuno and Y. Okada, Rev. Mod. Phys. {\bf 73}, 151 (2001).

\bibitem{PDG14} K. Olive {\it et al.} (Particle Data Group), Chin. Phys. C. {\bf 38} 090001 (2014).

\bibitem{zee} A.~Zee, Phys. Lett. B {\bf 93}, 389 (1980); Phys. Lett. B {\bf 95}, 461 (1980).

\bibitem{ATLAS} G.~Aad {\it et al.} [Atlas Collaboration], JHEP {\bf 1503}, 041 (2015).

\bibitem{CMS} S.~Chatrchyan {\it et al.} [CMS Collaboration], Eur. Phys. J. C {\bf 72}, 2189 (2012).

\bibitem{dbeta1} J. Barry, W. Rodejohann and H. Zhang, JHEP {\bf 1107}, 091 (2011).

\bibitem{dbeta2} C. Giunti and E. M. Zavanin, JHEP {\bf 1507}, 171 (2015).

\bibitem{vogelandbeacom} P. Vogel and J.F. Beacom, Phys. Rev. D {\bf 60}, 053003 (1999). We use the version of the cross sections that are corrected to order 1/M, as summarized by Eq. (18).  This reference is quoted as the cross section source for both LSND and KARMEN.

 \bibitem{fandc} G. J. Feldman and R. D. Cousins, Phys. Rev. D {\bf 57}, 3873 (1998).
  
  \bibitem{cems02} E.~D.~Church, K.~Eidel, G.~B.~Mills and M.~Steidl, Phys. Rev. D {\bf 66}, 013001 (2002).
  
 \bibitem{glw} C. Giunti, M. Laveder and W. Winter, Phys. Rev. D {\bf 80}, 073005 (2009).
 
 \bibitem{jsns2} M. Harada {\it et al.} (JSNS$^2$ Collaboration), arXiv:1601.01046.
 
 \bibitem{FS} Fumihiko Suekane, arXiv:1604.06190v2.  This reference reviews plans for the JSNS$^2$, OscSNS and KPipe experiments.
 
 \bibitem{IceCube} M. Aartsen {\it et al.} (ICECube PINGU) "Letter of Intent: The Precision IceCube Next Generation Upgrade (PINGU)" 1401.2046 (2014).
 

\end{thebibliography}
\end{document}